\documentclass[11pt]{article}
\usepackage{erice,epsfig}

\def\be{\begin{equation}}
\def\ee{\end{equation}}
\def\bea{\begin{eqnarray}}
\def\eea{\end{eqnarray}}

\def\beq{\begin{equation}}
\def\eeq{\end{equation}}

\def\be{\beta}
\def\ga{\gamma}
\def\de{\delta}
\def\ep{\epsilon}

\def\ka{\kappa}
\def\la{\lambda}

\def\si{\sigma}

\def\ps{\psi}
\def\om{\omega}
\def\Ga{\Gamma}

\def\mn{{\mu\nu}}

\def\fr#1#2{{{#1} \over {#2}}}

\def\half{{\textstyle{1\over 2}}}
\def\frac#1#2{{\textstyle{{#1}\over {#2}}}}

\def\lsim{\mathrel{\rlap{\lower4pt\hbox{\hskip1pt$\sim$}}
    \raise1pt\hbox{$<$}}}
\def\gsim{\mathrel{\rlap{\lower4pt\hbox{\hskip1pt$\sim$}}
    \raise1pt\hbox{$>$}}}
\def\sqr#1#2{{\vcenter{\vbox{\hrule height.#2pt
         \hbox{\vrule width.#2pt height#1pt \kern#1pt
         \vrule width.#2pt}
         \hrule height.#2pt}}}}

\begin{document}
\vspace*{4cm}
\title{ELECTROMAGNETIC TESTS OF LORENTZ AND CPT SYMMETRY}

\author{R.\ BLUHM}

\address{Physics Department, Colby College,
Waterville, ME 04901, USA}

\maketitle\abstracts{
A review is presented of some recent Lorentz and CPT tests in 
atomic and particle systems where the predominant
interactions are described by quantum electrodynamics.
A theoretical framework extending QED in the 
context of the standard model is used to
analyze these systems.
Experimental signatures of possible Lorentz
and CPT violation are investigated,
and bounds are discussed.}

\section{Introduction}

In recent years,
there has been a growing interest in testing Lorentz
and CPT symmetry.\cite{cpt02}
This is due to both theoretical developments
as well as improved experimental capabilities.
For example,
it has been shown that string theory can
lead to violations of CPT and Lorentz symmetry.\cite{kskp}
This is because strings are nonpointlike and have nonlocal interactions.
They can therefore evade the CPT theorem.
In particular,
there are mechanisms in string theory that can induce
spontaneous breaking of Lorentz and CPT symmetry.
It has also been shown that geometries with noncommutative
coordinates can arise naturally in string theory\cite{connes98}
and that Lorentz violation is intrinsic to noncommutative field
theories.\cite{chlkO01}

Experimental signals due to effects in these kinds of 
theories are expected at the Planck scale,
$M_{\rm Pl} = \sqrt{\hbar c/G} \simeq 10^{19}$ GeV,
where particle physics meets up with gravity.
This energy scale is inaccessible in accelerator experiments.
However,
a promising apprach has been to adopt Lorentz and CPT
violation as a candidate signal of new physics originating
from the Planck scale.
The idea is to search for effects that are heavily suppressed
at ordinary energies,
e.g.,
with suppression factors proportional to the ratio of a 
low-energy scale to the Planck scale.
Normally,
such signals would be unobservable.
However,
with a unique signal such as Lorentz or CPT violation
(which cannot be mimicked in conventional physics)
the opportunity arises to search for effects originating from the Planck scale.
This approach to testing Planck-scale physics has been aided  
by the development of a consistent theoretical framework
incorporating Lorentz and CPT violation in an extension of the 
standard model of particle physics.\cite{ck}
In the context of this framework,
it is possible to look for new signatures of Lorentz and CPT violation
in atomic and particle systems that might otherwise be overlooked.

Experiments in QED systems are particularly well suited to this approach
since they are often sensitive to extremely low energies.
Experiments in atomic physics are routinely sensitive
to small frequency shifts at the level of 1 mHz or less.
Expressing this as an energy shift in GeV
corresponds to a sensitivity of approximately
$4 \times 10^{-27}$ GeV.
Such a sensitivity is well within the range of energy one
might associate with suppression factors originating
from the Planck scale.
For example,
the fraction $m_p/M_{\rm Pl}$ multiplying the proton mass
yields an energy of approximately $10^{-19}$ GeV,
while for the electron the fraction $m_e/M_{\rm Pl}$ times
the electron mass is about $2.5 \times 10^{-26}$ GeV.

Some examples of QED systems that are highly sensitive
to Lorentz and CPT violation include experiments with
photons,\cite{cfj90,jk99,km01}
electrons,\cite{dehmelt99,mittleman99,bkr9798,bk00,eotwash}
muons,\cite{muonium01,muong01,bkl00}
protons,\cite{Hmaser,bkr99}
and
neutrons.\cite{dualmaser}
These examples include some of the classic tests of Lorentz and CPT symmetry,
such as $g-2$ experiments in Penning traps\cite{penningtests}
and atomic-clock comparisons -- the so-called 
Hughes-Drever experiments.\cite{cctests,kl99}
In addition to these examples involving leptons and baryons,
there are other experiments that provide bounds on
mesons.\cite{ckpv,mesons}

In the next section,
I begin with a brief review of Lorentz and CPT symmetry.
This includes a discussion of some of the theoretical 
ideas that have been put forward over the years for ways in which Lorentz 
symmetry and CPT might be violated in nature.
Different theoretical approaches to searching for Lorentz violation
are also described.
I then briefly review the standard-model extension.
It is the QED sector of this theory that is used
to investigate recent electromagnetic tests of 
Lorentz and CPT symmetry.
These tests are described in the subsequent sections,
with the photon and fermion sectors treated separately.
Several new Lorentz and CPT bounds are summarized.
Lastly,
some recent ideas involving possible tests of Lorentz and
CPT symmetry in a space satellite 
are presented.\cite{bklr02}

\section{Lorentz and CPT Symmetry}

It appears that nature is invariant under Lorentz symmetry and CPT.\cite{cptrev}
All physical interactions seem to be invariant under continuous Lorentz
transformations consisting of boosts and rotations and under the
combined discrete symmetry CPT
formed from the product of charge conjugation C,
parity P,
and time reversal T.
The CPT theorem links these symmetries.\cite{cptthm}
It states that (under mild technical assumptions) all local
relativistic field theories of point particles are symmetric under CPT.
A consequence of the CPT theorem is that particles and antiparticles should
have exactly equal lifetimes, masses, and magnetic moments.

Numerous experiments confirm Lorentz and CPT symmetry to
extremely high precision.
The Hughes-Drever type experiments are widely considered the
best tests of Lorentz symmetry.
These experiments place very stringent bounds on spatially 
anisotropic interactions.\cite{cctests}
The best CPT test listed by the Particle Data Group\cite{pdg}
compares the masses of neutral $K^0$ mesons with their antiparticles
and obtains a bound on their difference of a few parts in $10^{-19}$.

\subsection{Experimental Tests in QED Systems}

Many of the sharpest tests of CPT and Lorentz symmetry are made
in particle and atomic systems where the predominant interactions
are described by QED.
For example,
the Hughes-Drever type experiments typically compare two
clocks or high-precision magnetometers consisting of different atomic species.
The best CPT tests for leptons and baryons cited by the
Particle Data Group are made by atomic physicists working with Penning traps.
These experiments have obtained bounds of order $2 \times 10^{-12}$ on the  
relative difference in the $g$-factors of electrons and positrons
and of order $9 \times 10^{-11}$
on the relative difference in the charge-to-mass ratios
of protons and antiprotons.
In addition to these,
two proposed experiments at CERN intend to make high-precision
spectroscopic comparisons of trapped hydrogen and
antihydrogen.\cite{cernhbar} 
One possibility is to compare 1S-2S transitions
in hydrogen and antihydrogen.
These are forbidden transitions and can only occur as
two-photon transitions.
They have a small relative linewidth of approximately $10^{-15}$.
High precision comparisons of these and other transitions in
hydrogen and antihydrogen will yield sharp new CPT bounds.

It is interesting to note that of all the experiments testing 
Lorentz and CPT in matter it
is the atomic experiments which have the highest experimental precisions
(as opposed to sensitivity).
For example,
in neutral meson experiments quantities are measured with precisions
of approximately $10^{-3}$,
while in atomic experiments frequencies are typically measured with
precisions of $10^{-9}$ or better.
Nonetheless,
the CPT bound from the neutral meson experiments
is many orders of magnitude better than those 
from the atomic experiments.
It would therefore be desirable to understand the atomic experiments
better and to gain greater insight into their sensitivity.
Part of the difficulty in doing this stems from the fact that
these experiments all compare different physical quantities,
such as masses, $g$ factors, charge-to-mass ratios, and frequencies.
One way to find a more meaningful approach to making cross comparisons
would be to work within a common theoretical framework.

\subsection{Ideas for Violation}

A number of different ideas for violation of Lorentz or CPT symmetry 
have been put forward over the years.
In order to evade the CPT theorem one or more of the assumptions 
in the proof of the theorem must be disobeyed.
A sampling of some of the theoretical ideas that have been put forward
include the following:
nonlocal interactions,\cite{carr68}
infinite component fields,\cite{ot68}
a breakdown of quantum mechanics in gravity,\cite{hawk82}
and spontaneous Lorentz and CPT violation
occuring in the context of string theory.\cite{kskp}
It has also recently been shown that Lorentz violation is intrinsic 
to noncommutative field theories.\cite{chlkO01}

To investigate some of the experimental consequences of possible 
Lorentz or CPT violation,
a common approach has been to introduce phenomenological parameters.
Examples include the anisotropic inertial mass
parameters in the model of Cocconi and Salpeter,\cite{cs58}
the $\de$ parameter used in kaon physics,\cite{lw66}
and the TH$\ep\mu$ model which couples gravity and electromagnetism.\cite{ll73}
Another approach is to introduce specific types of lagrangian terms
that violate Lorentz or CPT symmetry.\cite{cfj90}
These approaches are straightforward and are largely model independent.
However,
they also have the disadvantage that their predictive ability
across different experiments is limited.
To make further progress,
one would want a consistent fundamental theory with Lorentz and CPT violation.
This would permit the calculation of phenomenological parameters
and the prediction of signals indicating symmetry violation.
No such realistic fundamental theory is known at this time.
However,
a candidate extension of the standard model incorporating
CPT and Lorentz violation does exist.

\subsection{The Standard-Model Extension}

An extension of the standard model incorporating Lorentz and
CPT violation has been developed by Kosteleck\'y and collaborators.
It provides a consistent theoretical framework that includes the
standard model (and SU(3)$\times$SU(2)$\times$U(1) gauge invariance)
and which allows for small violations of Lorentz and CPT symmetry.\cite{ck}
It is motivated in part from string theory and includes any 
low-energy effective theory that could
arise from spontaneous breaking of Lorentz symmetry.\cite{kskp}
The idea in this context is to assume the existence of a fundamental 
theory such as string theory in which Lorentz and CPT symmetry hold 
exactly but are spontaneously broken at low energy.
As in any theory with spontaneous symmetry breaking,
the symmetries become hidden at low energy.
The effective low-energy theory contains the standard model
as well as additional terms that could arise through the symmetry breaking process.
A viable realistic fundamental theory is not known at this time,
though higher dimensional theories such as string or M theory are promising candidates.  
A mechanism for spontaneous symmetry breaking
can be realized in string theory because
suitable Lorentz-tensor interactions can arise which destabilize
the vacuum and generate nonzero tensor vacuum expectation values.
It has been shown that any realistic noncommutative field theory
is equivalent to a subset of the standard-model extension.\cite{chlkO01}

Colladay and Kosteleck\'y have written down a general extension
of the standard model that could arise from spontaneous Lorentz
symmetry breaking of a more fundamental theory,
which maintains SU(3)$\times$SU(2)$\times$U(1) gauge invariance,
and is power-counting renormalizable.\cite{ck}
They have shown that the theory maintains many of
the other usual properties of the standard model besides Lorentz and CPT symmetry,
such as electroweak breaking, energy-momentum conservation,
the spin-statistics connection, and observer Lorentz covariance.
Issues related to the stability and causality of the standard-model
extension have been investigated as well.\cite{kl01}

\section{QED Extension}

To consider experiments involving electromagnetic interactions 
it suffices to restrict the standard-model extension to its QED sector.
The lagrangian describing electromagnetic interactions of a fermion
field $\ps$ of mass $m$ and charge $q = -|e|$ with photons $A^\mu$
can be written as
\beq
{\cal L} = {\cal L}_0 + {\cal L}_{\rm photon} + {\cal L}_{\rm fermion}
\quad .
\label{lag}
\eeq
Here,
${\cal L}_0$ is the usual QED lagrangian
in the absence of Lorentz and CPT violation,
\beq
{\cal L}_0 = i \bar \ps \ga^\mu D_\mu \ps - \bar \ps m \ps
-\fr 1 4 F_{\mu\nu} F^{\mu\nu} 
\quad ,
\label{qedlag}
\eeq
where $i D_\mu = i \partial_\mu - q A_\mu$,
$F_{\mu\nu} = \partial_\mu A_\nu - \partial_\nu A_\mu$,
and natural units with $\hbar = c = 1$ are used.
The Lorentz and CPT violating terms are
\beq
{\cal L}_{\rm photon} = 
\fr 1 2 (k_{AF})^\ka \ep_{\ka \la \mu \nu} A^{\la} F^{\mu \nu} 
- \fr 1 4 (k_F)_{\ka \la \mu \nu} F^{\ka \la} F^{\mu \nu}
\quad 
\label{photonlag}
\eeq
for the photon sector and 
\beq
{\cal L}_{\rm fermion} = - a_\mu \bar \ps \ga^\mu \ps
- b_\mu \bar \ps \ga _5 \ga^\mu \ps
- \half H_{\mu \nu} \bar \ps \si^{\mu \nu} \ps 
+ ic_{\mu \nu} \bar \ps \ga^\mu D^\nu \ps
+ id_{\mu \nu} \bar \ps \ga_5 \ga^\mu D^\nu \ps 
\quad 
\label{fermionlag}
\eeq
for the fermion sector.

Each of the additional terms involves a constant parameter.  
The terms involving the effective coupling constants
$a_\mu$, $b_\mu$, and  $(k_{AF})_\mu$ violate CPT,
while the terms with
$H_{\mu \nu}$, $c_{\mu \nu}$, $d_{\mu \nu}$, and 
$(k_F)_{\ka \la \mu \nu}$ preserve CPT.
All seven terms break Lorentz symmetry.\cite{note1}
The renormalizability of this theory has recently
been shown to hold to one loop.\cite{klp01}
The QED extension has also been used to study scattering cross
sections of electrons and positrons in the presence of
CPT and Lorentz violation.\cite{ck01}
In the following sections,
the photon and fermion sectors will be discussed separately.

\section{Photon Sector}

The extra terms in ${\cal L}_{\rm photon}$ lead to modifications
of Maxwell's equations and the energy density and dispersion
relations for photons.
A thorough discussion of these modifications
is given by Colladay and Kosteleck\'y.\cite{ck}
Here, I will briefly summarize some of the theoretical issues
and experimental bounds for these terms.
In many respects,
the theory for the photon sector
is analogous to electromagnetism in certain types of macroscopic media,
such as a crystal.
This results in the Lorentz and CPT violation
causing effects such as photon birefringence.
Bounds can therefore be obtained from experiments looking
at photons originating from cosmological sources.

The CPT-odd term involving $(k_{AF})_\mu$ gives rise to
negative-energy contributions,\cite{cfj90}
which would cause instabilities in the theory.
However,
this term is expected to vanish for theoretical reasons.\cite{ck}
It can be set to zero at tree level,
and then the question arises as to whether it acquires radiative
corrections from quantum loop corrections.
Remarkably,
the structure of the standard-model extension leads to
an anomaly cancelation mechanism that preserves the
vanishing of $(k_{AF})_\mu$ at the one-loop level.\cite{ck,jk99}
In addition to these theoretical constraints,
very sharp bounds on $(k_{AF})_\mu$ at the level of 
$10^{-42}$ GeV can be obtained from cosmological birefringence experiments.\cite{cfj90}
For these reasons,
$(k_{AF})_\mu$ will be assumed to vanish in the following sections.

The CPT-even term involving $(k_F)_{\ka \la \mu \nu}$ leads
to positive energy contributions.
There are no theoretical reasons to expect that it vanishes.
This term has 19 independent real components.
Their contributions have been shown to
result in a wavelength dependence in the relative phase difference between
the photon polarizations.\cite{km01}
This gives rise to a new method of extracting bounds 
in spectropolarimetry of cosmological sources.
A recent survey of different sources\cite{km01}
results in bounds at the level of $10^{-32}$ on many of the 
contributions from the $(k_F)_{\ka \la \mu \nu}$ term.

\section{Fermion Sector}

The effects of Lorentz and CPT violation in matter are
controlled by the term ${\cal L}_{\rm fermion}$ in the lagrangian.
It involves the five effective coupling constants
$a_\mu$, $b_\mu$, $H_{\mu \nu}$, $c_{\mu \nu}$, and $d_{\mu \nu}$,
which are all assumed to be small.
It is these terms that cause leading-order corrections in
QED systems involving matter.
They effectively give rise to a modified structure for
the mass and gamma matrices in the Dirac equation,
\beq
( i \Ga^\mu D_\mu - M) \ps = 0
\quad ,
\label{dirac}
\eeq
where
\beq
\Ga_\nu = \ga_\nu + c_\mn \ga^\mu + d_\mn \ga_5 \ga^\mu
\quad ,
\label{Gam}
\eeq
and
\beq
M = m + a_\mu \ga^\mu + b_\mu \ga_5 \ga^\mu
   + \half H_\mn \si^\mn
\quad .
\label{M}
\eeq
Additional interactions involving the photon Lorentz-violation parameters
($(k_{AF})_\mu$ or $(k_F)_{\ka \la \mu \nu}$)
coupling to matter through the photon propagator will be of sub-leading order
and can be ignored.
The leading-order corrections can then be found using relativistic
quantum mechanics in a perturbative treatment.

In the last several years,
a number of experiments in QED systems have been performed that
have resulted in sharp new bounds on Lorentz and CPT violation.
These bounds are typically expressed in terms of the parameters
$a_\mu$, $b_\mu$, $c_{\mu \nu}$, $d_{\mu \nu}$, and $H_{\mu \nu}$.
This permits a straightforward way of making comparisons across different
types of experiments and avoids problems that can arise
when different physical quantities
($g$ factors, charge-to-mass ratios, masses, frequencies, etc.)
are used in different experiments.
A thorough investigation of possible CPT and Lorentz violation
must look at as many different particle sectors as possible,
since each different particle sector in the QED extension
has a set of Lorentz-violating parameters that are independent.
The parameters of the different sectors are distinguished
in the following using superscript labels.
Recent experiments have obtained bounds on parameters for the
electron,\cite{dehmelt99,mittleman99,bkr9798,bk00,eotwash}
muon,\cite{muonium01,muong01,bkl00}
proton,\cite{Hmaser,bkr99}
and
neutron.\cite{dualmaser}

Before discussing these recent experiments individually,
it is useful to examine some of the more general
results that have emerged from these investigations.
First, the sharp distinction between what are considered Lorentz tests
and CPT tests has been greatly diminished.
Experiments traditionally viewed as CPT tests are
also sensitive to Lorentz symmetry and vice versa.
In particular,
it has been demonstrated that it is possible to test for CPT 
violation in experiments with particles alone.
This has opened up a whole new arena of CPT tests.
A second general feature is that
the sensitivity to CPT and Lorentz violation in these experiments
stems primarily from their
ability to detect very small anomalous energy shifts.
While many of the experiments were originally designed to
measure specific quantities,
such as differences in $g$ factors or
charge-to-mass ratios of particles and antiparticles,
it is now recognized that they are most effective as
CPT and Lorentz tests when all of the energy levels in the system
are investigated for possible anomalous shifts.
Because of this,
several new signatures of CPT and Lorentz violation have been
investigated in recent years that were previously overlooked.
Examples are given in the following sections.
Finally,
another common feature of these experiments is that they
all have sensitivity to the Planck scale.

\subsection{Penning-Trap Experiments}

The aim of the original experiments with Penning traps was to
make high-precision comparisons of the $g$ factors and charge-to-mass ratios of
particles and antiparticles confined within the trap.\cite{penningtests}
This was obtained through measurements of the
anomaly frequency $\om_a$ and the cyclotron frequency $\om_c$.
For example,
$g-2=2\om_a/\om_c$.
The frequencies were typically measured to $\sim 10^{-9}$ for the electron,
which determines $g$ to $\sim 10^{-12}$.
In computing these ratios it was not necessary to keep
track of the times when $\om_a$ and $\om_c$ were measured.
More recently,
however,
additional signals of possible CPT and Lorentz violation
in this system have been found,
which has led to two new tests being performed.

The first was a reanalysis performed by Dehmelt's group of existing
data for electrons and positrons in a Penning trap.\cite{dehmelt99}
The goal was to search for an instantaneous difference in the
anomaly frequencies of electrons and positrons,
which can be nonzero when CPT and Lorentz symmetry are broken.
(In contrast the leading-order instantaneous cyclotron frequencies 
remain equal).
The original measurements of $g-2$
did not involve looking for possible instantaneous variations in $\om_a$.
Instead,
the ratio $\om_a/\om_c$ was obtained using averaged values.
The new analysis is especially relevant because it can be shown that
the CPT-violating corrections to the anomaly frequency
$\om_a$ can occur even though the $g$ factor remains unchanged.
The new bound found by Dehmelt's group can be expressed in terms of 
the parameter $b^e_3$,
which is the component of $b^e_\mu$ along the quantization
axis in the laboratory frame.
They obtained $|b^e_3| \lsim 3 \times 10^{-25}$ GeV.

A second new test of CPT and Lorentz violation in the electron
sector has been made using only data for the electron.\cite{mittleman99}
Here,
the idea is that the Lorentz and CPT-violating interactions depend on
the orientation of the quantization axis in the laboratory frame,
which changes as the Earth turns on its axis.
As a result,
both the cyclotron and anomaly frequencies have small corrections which
cause them to exhibit sidereal time variations.
Such a signal can be measured using just electrons,
which eliminates the need for comparison with positrons.
The bounds in this case are given with respect to a
nonrotating coordinate system such as celestial equatorial coordinates.
The interactions involve a combination of laboratory-frame components
that couple to the electron spin.
The combination is denoted as
$\tilde b_3^e  \equiv b_3^e - m d_{30}^e - H_{12}^e$.
The bound can be expressed in terms of components $X$, $Y$, $Z$ 
in the nonrotating frame.
It is given as
$|\tilde b_J^e| \lsim 5 \times 10^{-25} {\rm GeV}$ for $J=X,Y$.

\subsection{Clock-Comparison Experiments}

The Hughes-Drever experiments
are classic tests of Lorentz invariance.\cite{cctests}
These experiments look for relative changes between two atomic ``clock''
frequencies as the Earth rotates.
The ``clock'' frequencies are typically atomic hyperfine or Zeeman transitions.
In a 1995 experiment, 
very sharp bounds at leading-order for the proton, neutron and
electron were obtained in the experiment of Berglund {\it et al}.
These were
$\tilde b_J^p \simeq 10^{-27}$ GeV,
$\tilde b_J^n \simeq 10^{-30}$ GeV,
and $\tilde b_J^e \simeq 10^{-27}$ GeV for $J=X,Y$.

More recently,
several new clock-comparison tests have been performed
or are in the planning stages.
For example,
Bear {\it et al.} have used a two-species noble-gas maser to
test for Lorentz and CPT violation in the neutron sector.\cite{dualmaser}
They obtained a bound
$|\tilde b_J^n| \lsim 10^{-31} {\rm GeV}$ for $J=X,Y$.
This is currently the best bound for the neutron sector.
As sharp as these bounds are,
however,
it should be kept in mind that certain assumptions about the nuclear
configurations must be made to obtain them.
For this reason,
these bounds should be viewed as accurate to within one or two
orders of magnitude.
To obtain cleaner bounds it is necessary to consider
simpler atoms or to perform more precise nuclear modeling.

\subsection{Hydrogen-Antihydrogen Experiments}

Hydrogen atoms have the simplest nuclear structure.
Two experiments are being planned at CERN which will
make high-precision spectroscopic measurements of the 1S-2S
transitions in hydrogen and antihydrogen.
These are forbidden two-photon transitions with a relative linewidth
of approximately $10^{-15}$.
The magnetic field plays an important role
in the sensitivity of these transition to Lorentz and CPT breaking.
For example,
in free hydrogen in the absence of a magnetic field,
the 1S and 2S levels shift by the same amount at leading order,
and there are no leading-order corrections to the 1S-2S transition.
However,
in a magnetic trap
there are fields that mix the different spin states in the
four hyperfine levels.
Since the Lorentz-violating couplings are spin-dependent,
there will be sensitivity at leading order
to Lorentz and CPT violation in comparisons of 1S-2S transitions in
trapped hydrogen and antihydrogen.

An alternative to 1S-2S transitions is to consider 
measurements of ground-state Zeeman
hyperfine transitions in hydrogen alone.
It has been shown that these transitions in a hydrogen maser
are sensitive to leading-order Lorentz-violating effects.
Measurements of these transitions have recently 
been made using a double-resonance
technique.\cite{Hmaser}
They yield new bounds for the electron and proton.
The bound for the proton is $|\tilde b_J^p| \lsim 10^{-27}$ GeV.
Due to the simplicity of the hydrogen nucleus,
this is an extremely clean bound.
It is currently the best
Lorentz and CPT symmetry for the proton.

\subsection{Spin-Polarized Matter}

A recent experiment at the University of Washington uses a spin-polarized
torsion pendulum to achieve high sensitivity to
Lorentz violation in the electron sector.\cite{eotwash}
Its sensitivity stems from the combined effect of a large number
of aligned electron spins.
The experiment uses stacked toroidal magnets that have a net
electron spin $S \simeq 8 \times 10^{22}$,
but which have a negligible magnetic field.
The pendulum is suspended on a turntable and a time-varying
harmonic signal is sought.
An analysis of this system reveals that in addition to a signal with the
period of the rotating turntable,
the effects of Lorentz and CPT violation induce additional
time variations with a sidereal period caused by Earth's rotation.
The group at the University of Washington has analyzed their data
and has obtained a bound on the electron parameters
equal to $|\tilde b_J^e| \lsim 10^{-29}$ GeV for $J=X,Y$ and
$|\tilde b_Z^e| \lsim 10^{-28}$ GeV.\cite{eotwash}
These are currently the best Lorentz and CPT bounds for the electron.

\subsection{Muon Experiments}

Experiments with muons involve second-generation leptons.
They provide independent Lorentz and CPT tests.
There are several different kinds of experiments with muons
that are currently being conducted,
including muonium experiments\cite{muonium01}
and $g-2$ experiments with muons at Brookhaven.\cite{muong01}
In muonium,
the experiments measuring the frequencies
of ground-state Zeeman hyperfine transitions
in a strong magnetic field have the best sensitivity
to Lorentz and CPT violation.
A recent analysis has looked for sidereal time variations
in these transitions.
A bound at a level of $| \tilde b^\mu_J| \le 5 \times 10^{-22}$ GeV
has been obtained.\cite{muonium01}
The $g-2$ experiments with positive muons
are relativistic with ``magic'' boost parameter $\de = 29.3$.
Bounds on Lorentz-violation parameters should be attainable in these
experiments at a level of $10^{-25}$ GeV.
These experiments are currently underway at Brookhaven and their results
should be forthcoming in the near future.

\section{Clock-Comparison Experiments in Space}

In summary,
five new sets of Lorentz and CPT bounds have been obtained 
in recent years for the 
electron, proton, neutron, and muon.
The leading-order bounds are summarized in Table 1.
All of these limits are within the range of sensitivity associated
with suppression factors arising from the Planck scale.
However,
as sharp as these bounds are,
there is still room for improvement,
and it is likely that the next few years will continue to provide
increasingly sharp new tests of Lorentz and CPT symmetry in QED systems.
In particular,
it should be possible to obtain bounds on many of the parameters
that do not appear in Table 1.

\begin{table}[t]
\begin{center}
\noindent
\renewcommand{\arraystretch}{1.2}
\begin{tabular}{|c|c|c|c|}
\hline\hline
Expt & Sector & Params ($J=X,Y)$ & Bound (GeV)
\\
\hline\hline
Penning Trap & electron & $\tilde b_J^e$ & $5 \times 10^{-25}$ \\[2mm]
\cline{1-4}
Hg-Cs clock  & electron & $\tilde b_J^e$ & $\sim 10^{-27}$ \\[2mm]
\cline{2-4}
comparison & proton & $\tilde b_J^p$ & $\sim 10^{-27}$ \\[2mm]
\cline{2-4}
 & neutron & $\tilde b_J^n$ & $\sim 10^{-30}$ \\[2mm]
\cline{1-4}
He-Xe dual maser & neutron & $\tilde b_J^n$ & $\sim 10^{-31}$ \\[2mm]
\cline{1-4}
H maser & electron & $\tilde b_J^e$ & $10^{-27}$ \\[2mm]
\cline{2-4}
 & proton & $\tilde b_J^p$ & $10^{-27}$ \\[2mm]
\cline{1-4}
Spin Pendulum & electron & $\tilde b_J^e$ & $10^{-29}$ \\[2mm]
&& $\tilde b_Z^e$ & $10^{-28}$ \\[2mm]
\cline{1-4}
Muonium & muon & $\tilde b_J^\mu$ & $2 \times 10^{-23}$ \\[2mm]
\cline{1-4}
Muon $g-2$ & muon & $\mathaccent 20 b_J^\mu$ & $5 \times 10^{-25}$ \\[2mm]
&&& {\rm (estimated)} \\[2mm]
\hline
\hline
\end{tabular}
\renewcommand{\arraystretch}{1.0}
\caption{Summary of leading-order bounds.}
\vspace{0.2cm}
\end{center}
\end{table}

One promising approach is to conduct atomic clock-comparison tests
in a space satellite.\cite{bklr02}
These will have several advantages over traditional
ground-based experiments, 
which are typically insensitive to the
direction $Z$ of Earth's axis and ignore boost
effects associated with timelike directions.
For example,
a clock-comparison experiment conducted aboard 
the International Space Station (ISS)
would be in a laboratory frame that is both rotating and boosted.
It would therefore immediately gain sensitivity to
both the $Z$ and timelike directions.
This would more than triple the number of Lorentz-violation
parameters that are accessible in a clock-comparison experiment.
Since there are several missions already planned for
the ISS which will compare Cs and Rb atomic clocks and H masers,
the opportunity to perform these new Lorentz and CPT
tests is quite real.
Another advantage of an experiment aboard the ISS is
that the time needed to acquire data would be greatly
reduced (by approximately a factor of 16).
In addition,
new types of signals would emerge that have no
analogue in traditional Earth-based experiments.
The combination of these advantages should result in
substantially improved limits on Lorentz and CPT violation.

\section*{Acknowledgments}
I would like to acknowledge my collaborators
Alan Kosteleck\'y, Charles Lane, and Neil Russell.
This work is supported in part by the National
Science Foundation under grant number PHY-9801869.



\section*{References}
\begin{thebibliography}{99}

\bibitem{cpt02}
See, for example,
V.A. Kosteleck\'y, ed.,
{\it CPT and Lorentz Symmetry II}
(World Scientific, Singapore, 2002).

\bibitem{kskp}
V.A.\ Kosteleck\'y and S.\ Samuel,
Phys.\ Rev.\ D {\bf 39} (1989) 683;
{\it ibid.}
{\bf 40} (1989) 1886;
Phys.\ Rev.\ Lett.\ {\bf 63} (1989) 224;
{\it ibid.}
{\bf 66} (1991) 1811;
V.A.\ Kosteleck\'y and R.\ Potting,
Nucl.\ Phys.\ B {\bf 359} (1991) 545;
Phys.\ Lett.\ B {\bf 381} (1996) 89.

\bibitem{connes98}
A. Connes, M. Douglas, and A. Schwarz,
JHEP {\bf 02} (1998) 003.

\bibitem{chlkO01}
S.M.\ Carroll, J.A.\ Harvey, V.A.\ Kosteleck\'y, C.D.\ Lane, and T.\ Okamoto,
Phys.\ Rev.\ Lett.\ {\bf 87}, 141601 (2001).

\bibitem{ck}
D.\ Colladay and V.A.\ Kosteleck\'y,
Phys.\ Rev.\ D {\bf 55} (1997) 6760;
Phys.\ Rev.\ D {\bf 58}, 116002 (1998).

\bibitem{cfj90}
S.M.\ Carroll, G.B.\ Field, and R.\ Jackiw,
Phys.\ Rev.\ D {\bf 41} (1990) 1231.

\bibitem{jk99}
R.\ Jackiw and V.A.\ Kosteleck\'y,
Phys.\ Rev.\ Lett.\ {\bf 82} (1999) 3572.

\bibitem{km01}
V.A.\ Kosteleck\'y and M.\ Mewes,
Phys.\ Rev.\ Lett.\ {\bf 87}, 251304 (2001).

\bibitem{dehmelt99}
H.G.\ Dehmelt
{\it et al.},
Phys.\ Rev.\ Lett.\ {\bf 83} (1999) 4694.

\bibitem{mittleman99}
R.\ Mittleman
{\it et al.},
Phys.\ Rev.\ Lett.\ {\bf 83} (1999) 2116.

\bibitem{bkr9798}
R.\ Bluhm, V.A.\ Kosteleck\'y and N.\ Russell,
Phys.\ Rev.\ Lett.\ {\bf 79} (1997) 1432;
Phys.\ Rev.\ D {\bf 57} (1998) 3932.

\bibitem{bk00}
R.\ Bluhm and V.A.\ Kosteleck\'y,
Phys. Rev. Lett. {\bf 84} (2000) 1381.

\bibitem{eotwash}
B.R.\ Heckel
{\it et al.},
to appear\cite{cpt02}.

\bibitem{muonium01}
V.W.\ Hughes {\it et al.},
Phys. Rev. Lett. {\bf 87}, 111804 (2001).

\bibitem{muong01}
H.N.\ Brown {\it et al.},
Phys. Rev. Lett. {\bf 86} (2001) 2227.

\bibitem{bkl00}
R. Bluhm, V.A.\ Kosteleck\'y and C.D. Lane,
Phys. Rev. Lett. {\bf 84} (2000) 1098.

\bibitem{Hmaser}
D.F.\ Phillips
{\it et al.},
Phys. Rev. D {\bf 63}, 111101 (2001).

\bibitem{bkr99}
R.\ Bluhm, V.A.\ Kosteleck\'y and N.\ Russell,
Phys.\ Rev.\ Lett.\ {\bf 82} (1999) 2254.

\bibitem{dualmaser}
D.\ Bear {\it et al.},
Phys.\ Rev.\ Lett.\ {\bf 85} (2000) 5038.

\bibitem{penningtests}
P.B.\ Schwinberg, R.S.\ Van Dyck, Jr., and H.G.\ Dehmelt,
Phys.\ Rev.\ D {\bf 34} (1986) 722;
R.S.\ Van Dyck, Jr., P.B.\ Schwinberg, and H.G.\ Dehmelt,
Phys.\ Rev.\ Lett.\ {\bf 59} (1987) 26;
G.\ Gabrielse
{\it et al.},
Phys.\ Rev.\ Lett.\ {\bf 82} (1999) 3198.

\bibitem{cctests}
V.W.\ Hughes, H.G.\ Robinson, and V.\ Beltran-Lopez,
Phys.\ Rev.\ Lett.\ {\bf 4} (1960) 342;
R.W.P.\ Drever,
Philos.\ Mag.\ {\bf 6} (1961) 683;
J.D.\ Prestage
{\it et al.},
Phys.\ Rev.\ Lett.\ {\bf 54} (1985) 2387;
S.K.\ Lamoreaux
{\it et al.},
Phys.\ Rev.\ A {\bf 39} (1989) 1082;
T.E.\ Chupp
{\it et al.},
Phys.\ Rev.\ Lett.\ {\bf 63} (1989) 1541;
C.J.\ Berglund
{\it et al.},
Phys.\ Rev.\ Lett.\ {\bf 75} (1995) 1879.

\bibitem{kl99}
V.A.\ Kosteleck\'y and C.D. Lane,
Phys. Rev. D {\bf 60}, 116010 (1999).

\bibitem{ckpv}
V.A.\ Kosteleck\'y and R.\ Potting,
Phys.\ Rev.\ D {\bf 51} (1995) 3923;
D.\ Colladay and V.A.\ Kosteleck\'y,
Phys.\ Lett.\ B {\bf 344} (1995) 259;
Phys.\ Rev.\ D {\bf 52} (1995) 6224;
V.A.\ Kosteleck\'y and R.\ Van Kooten,
Phys.\ Rev.\ D {\bf 54} (1996) 5585;
V.A.\ Kosteleck\'y,
Phys.\ Rev.\ Lett.\ {\bf 80} (1998) 1818;
Phys.\ Rev.\ D {\bf 61}, 016002 (2000);
{\bf 64}, 076001 (2001).
N.\ Isgur
{\it et al.},
Phys.\ Lett.\ B {\bf 515}, 333 (2001).

\bibitem{mesons}
B.\ Schwingenheuer
{\it et al.},
Phys.\ Rev.\ Lett.\ {\bf 74} (1995) 4376;
L.K.\ Gibbons
{\it et al.},
Phys.\ Rev.\ D {\bf 55} (1997) 6625;
OPAL Collaboration,
R.\ Ackerstaff {\it et al.},
Z.\ Phys.\ C {\bf 76} (1997) 401;
DELPHI Collaboration,
M.\ Feindt
{\it et al.},
preprint DELPHI 97-98 CONF 80 (July 1997).

\bibitem{bklr02}
R.\ Bluhm, V.A.\ Kosteleck\'y, C.D. Lane, and N.\ Russell,
to appear (hep-ph/0111141).

\bibitem{cptrev}
For reviews of CPT, see
R.F. Streater and A.S. Wightman,
{\it PCT, Spin, and Statistics and All That},
Benjamin Cummings, Reading, 1964;
R. G. Sachs,
{\it The Physics of Time Reversal},
University of Chicago, Chicago, 1987.

\bibitem{cptthm}
J. Schwinger,
Phys. Rev. {\bf 82} (1951) 914;
J.S. Bell,
Birmingham University thesis (1954);
Proc. Roy. Soc. (London) {\bf A 231} (1955) 479;
G. L\"uders,
Det. Kong. Danske Videnskabernes Selskab Mat.fysiske Meddelelser {\bf 28}, No. 5 (1954);
Ann. Phys. (N.Y.) {\bf 2} (1957) 1;
W. Pauli,
in W. Pauli, ed.,
{\it Neils Bohr and the Development of Physics},
McGraw-Hill, New York, 1955, p. 30.

\bibitem{pdg}
J. Bartels {\it et al.},
Review of Particle Properties,
Eur.\ Phys.\ J.\ C {\bf 15} (2000) 1.

\bibitem{cernhbar}
B. Brown {\it et al.},
Nucl. Phys. B (Proc. Suppl.) {\bf 56A} (1997) 326;
M.H. Holzscheiter {\it et al.},
Nucl. Phys. B (Proc. Suppl.) {\bf 56A} (1997) 336.

\bibitem{carr68}
P. Carruthers,
Phys. Lett. B {\bf 26} (1968) 158.

\bibitem{ot68}
A.I. Oksak and I.T. Todorov,
Commun. Math. Phys. {\bf 11} (1968) 125.

\bibitem{hawk82}
S. Hawking,
Commun. Math. Phys. {\bf 87} (1982) 395.

\bibitem{cs58}
G. Cocconi and E. Salpeter,
Nuovo Cimento {\bf 10} (1958) 646.

\bibitem{lw66}
See for example,
T.D. Lee and C.S. Wu,
Annu. Rev. Nucl. Sci {\bf 16} (1966) 511.

\bibitem{ll73}
A.P. Lightman and D.L. Lee,
Phys. Rev. D {\bf 8} (1973) 364.

\bibitem{kl01}
V.A.\ Kosteleck\'y and R.\ Lehnert,
Phys.\ Rev.\ D {\bf 63} (2001) 065008.

\bibitem{note1}
Additional fermion terms invloving parameters $e_\mu$, $f_\mu$,
and $g_{\la\mu\nu}$ are ignored.
Gauge invariance excludes these terms in the
standard-model extension.
However,
they could be generated effectively due to strong binding
in a nucleus and are included in a more general discussion.\cite{kl99}

\bibitem{klp01}
V.A.\ Kosteleck\'y, C.D.\ Lane and A.G.M.\ Pickering,
Phys.\ Rev.\ D, in press (hep-th/0111123).

\bibitem{ck01}
D.\ Colladay and V.A.\ Kosteleck\'y,
Phys.\ Lett.\ B {\bf 511}, 209 (2001).



\end{thebibliography}
\end{document}